\documentclass[%
twocolumn,
showpacs,
nofootinbib, nobibnotes,
amsmath,amssymb, aps,
pra,
floatfix,
]{revtex4}
\usepackage{graphicx}
\usepackage{placeins}
\usepackage{dcolumn}
\usepackage{bm}
\usepackage{hyperref}
\usepackage{natbib}
\usepackage{subfigure}
\usepackage{color}
\usepackage{amsmath}

\begin{document}

\preprint{APS/123-QED}

\title{On-Chip Photonic Transistor based on the Spike Synchronization in Circuit QED}

%

\author{Yusuf G\"ul}
\email{yusufgul.josephrose@gmail.com.}  \affiliation{Department of Physics, Bo\u{g}azi\c{c}i
University 34342 Bebek, Istanbul, Turkey} \affiliation{Momentum Research and Development, METU Technopolis 06800 \c{C}ankaya, Ankara, Turkey}

%
\date{\today}

\begin{abstract}
We consider the single photon transistor in coupled cavity system of resonators interacting with multilevel superconducting artificial atom simultaneously. Effective single mode transformation is used for the diagonalization of the hamiltonian and impedance matching in terms of the normal modes. Storage and transmission of the incident field are described by the interactions between the cavities controlling the atomic transitions of lowest lying states. Rabi splitting of vacuum induced multiphoton transitions is considered in input/output relations by the quadrature operators in the absence of the input field. Second order coherence functions are employed to investigate the photon blockade and delocalization-localization transitions of cavity fields. Spontaneous virtual photon conversion into real photons is investigated in localized and oscillating regimes. Reflection and transmission of cavity output fields are investigated in the presence of the multilevel transitions. Accumulation and firing of the reflected and transmitted fields are used to investigate the synchronization of the bunching spike train of transmitted field  and population imbalance of cavity fields. In the presence of single photon gate field, gain enhancement is explained for transmitted regime.
\end{abstract}



\maketitle
\section{Introduction}
Atom photon interactions lies at the heart of the emergence of the quantum optical devices templated by the on-chip lab simulation in microwave frequency quantum networks \cite{Feynman1982,lewenstein2006,Lukin2003,houck2012}. Cavity/Circuit QED leads to the implementation
of quantum information protocols of light-matter coupling in both the strong and ultrastrong regime\cite{Blais2004,Bourassa2009,abdulmalikov2010,peropadre2013}. Lattice arrays of coupled cavities mimicks the  collective behavior of many body coupled systems in quantum phase transitions\cite{greentree2006,hartman2008,schmidt2013,nissen2012}. Coupled resonator systems of cavities appears as the test bed for photon blockade and localization-delocalization transition in both strong and ultrastrong coupling regimes \cite{Schmidt2010,Didier2011,Ridolfo2012}.

Jahn-Teller (JT)\cite{kaplan1995,popovic2000} system as a solid state counterpart of the coupled cavity system is investigated in cavity QED \cite{larson2008}. In terms of two superconducting
LC resonators coupled with a common two-level artificial atom is used as a simulator in the ultrastrong coupling regime\cite{Dereli2012}. Two-frequency JT system can be switched to an effective single mode from the two-mode model by tuning the hopping term between the resonator in different coupling regimes \cite{O'Brien1972,O'Brien1983}. As an electrical analog platform to
simulate the nonlinearities \cite{Lefevre-Seguin1992,Bertet2005-1,Bertet2005-2,Reuther2011}, photon blockade and localization-delocalization transitions is investigated in coupled Superconducting Quantum Interference Devices
(SQUIDS). Population imbalance between coupled resonators is used to synchronize the qubit dephasing in networked JT system  wired up to outer circuitry on chip\cite{Gul2016}.

In analogy with the electronic counterpart, strong atom field interactions give rise to construction
of the single photon transistor based on the transmission of a strong coherent probe controlled by the
gate field\cite{Chang2007}. In this scheme, nonlinear two-photon switch is used to manipulate the propagating surface plasmon excitations in nanowires and the three-level atomic system \cite{Chen2013}. While the transmission of the signal photon is conditioned on the impedance matching in terms of the presence or absence of the single photon in the gate field, the storage process is described by the splitting of the incoming field due to the scattering by the emitter \cite{Shen2005,Shen2007,Shen2009}. Impedance matching in $\Lambda$ -type three-level systems, makes it possible to implement the absorption and switching in superconducting qubit resonator interactions \cite{Koshino2013,Inamato2014}.

Based on the single photon non-linearities, it is possible to propose different schemes of the resonator and artificial atom configurations in strong and ultrastrong coupling regimes. Two transmon qubits are used to couple the resonators on which photons are propagating and forming localized scattering center controlling the blockade and transmission of the incident photon\cite{Neumeier2013}. When the single artificial atom is three level ladder $\Xi$ system, single photon transistor is realised by reflected or transmitted photons are conditioned on the flipping of the atom\cite{Manzoni2014}. On the other hand,  two separate cavities can be wired up to construct single photon transistor by making the each cavity is coupled with the upper or lower transitions of a three level artificial atom with separate multiphoton input/output \cite{Kyriienko2016,Schoelkopf2008}. Moreover, coupled cavity system interacting with a qubit can be described in terms of the coordinates of the second auxiliary cavity effectively. Then, interactions between cavities and atomic transitions can be turned into the dark state tuned by the hopping term after eliminating the transitions of the multilevel atomic states\cite{Liu2014}. Input-output relations in the ultrastrong
coupling regime makes the multiphoton transitions possible in three-level cascaded atomic system. Conversion of the virtual photons into real photons is succeeded
by suitable designs of the Hamiltonians of the cascaded three-level atoms supporting the dark states \cite{Garziano2013,Stassi2013,Huang2014,Garziano2015}. In the presence of atomic transitions, Waveguide-QED systems reveal the importance of the correlation functions in describing the output cavity fields by relating the scattering theory and input output formalism of single and few photon transportation \cite{Fan2010,Shi2013,Kolchin2011}. Other than the superconducting systems, quantum optical transistors has also been proposed in the single-photon level \cite{Gorniaczyk2014,Tiarks2014,Hwang2009,Hu2016}

We consider the coupled cavity scheme of effective two-frequency JT system in which, single photon transistor is mimicked by the storage and transmission of the single photon in $\Lambda$-system of artificial atom\cite{Chang2007}. Interactions between the cavities is described by the hoping term corresponding to the atomic transitions in the lowest lying states of the artificial atom \cite{Liu2014}. Input/output formalism is employed to describe the extracavity emission in terms of the quadrature operators in the absence of the input field\cite{Garziano2013,Stassi2013,Huang2014,Garziano2015}. Two frequency JT hamiltonian is described as the interaction of the bright and dark polaritons and diagonalized in terms of the upper and lower polariton components. Splitting of the polaritons modes is investigated by the second order coherence functions.

This paper is organized as follows. In Sec.II we introduce the
coupled resonator model interacting with the multilevel atom and protocol for the single photon transistor. The results and discussions are presented in
Sec.III. Finally, we give conclusions in Sec. IV.

\section{Model and Protocol} \label{sec:model}


\subsection{Circuit Layout and Protocol}

Realization of our single photon protocol relies on the propagation of  “signal” photons
dependent on photon number in a “gate” field. Splitting of the incident field and the impedance matching are two essential
steps in single photon transistor schemes of coupled resonator systems. While the splitting of the incoming field leads to the
storage process, impedance matching is responsible for the perfect transmission and reflection of the signal field controlled by the gate field. \cite{Chang2007,Manzoni2014,Kyriienko2016}
\begin{figure}[h]
\begin{center}
\subfigure[\hspace{0.001cm}]{\label{fig:1a}
\includegraphics[width=0.4\textwidth]{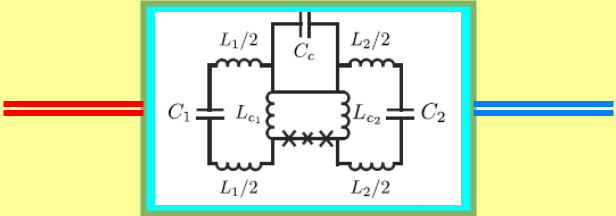}}
\subfigure[\hspace{0.001cm}]{\label{fig:1a}
\includegraphics[width=0.4\textwidth]{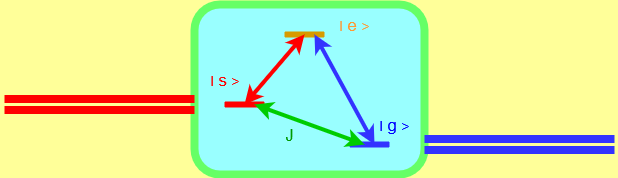}}
\subfigure[\hspace{0.001cm}]{\label{fig:1a}
\includegraphics[width=0.4\textwidth]{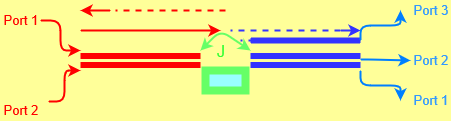}}
\caption{\label{fig3}(Color online)(a) Circuit model of two-frequency coupled resonator system.
A 3-Josephson junction
flux qubit capacitively coupled to LC resonators via coupling inductances $L_{c1;c2}$ of individual
inductors. The resonator are interacting with each other via capacitor $C_{c}$ which determines hopping parameter J together with individual capacitances $C_{1,2}$ and frequencies $\omega_{1,2}$. Two resonators further combined with microwave waveguide resonators or transmission line resonator (red and blue double lines) as input/output ports. (b) Coupled system of two cavities which play the role of the resonators and three level superconducting artificial atom in $\Lambda$ structure. Atomic energy levels resides in separate cavities. The metastable state $|s\rangle$ and ground state $|g\rangle$ which are localized in left (red) and right (blue) cavities are further coupled to each other by photon hopping controlled by $J$. Interaction strength of $J$ is determined by the coupling capacitor
$C_{c}$ in the physical circuit. The storage process takes place by triggering a spin
flip from $|g\rangle$ to $|s\rangle$ governed by the photonic nonlinearities with parameter $J$ and controls the reflection and transmission of cavity field. (c) Embedding of On-Chip cavity-artificial superconducting qubit system into microwave frequency waveguide or transmission line resonators serving as outer detection circuitry. Spontaneous virtual photon conversion into real photon can be detected by the input/output cavity ports.
Transmission (dashed blue arrow) and reflection (dashed red arrow) of the cavity fields (red arrow) are tuned by the hopping parameter $J$.}
\end{center}
\end{figure}

Our physical system consists of the lumped element LC resonators in Fig.1.a. The capacitive coupling between
resonators is mediated by $C_{c}$. A 3-phase Josephson junction flux qubit is used as an artificial superconducting atom simultaneously
coupled to each resonator through coupling
inductances $L_{c_{1},c_{2}}$ of individual inductors \cite{Pol2010,Dereli2012,peropadre2013}. The coupling strength between resonators is tuned by hopping parameter $J=C_{c}\sqrt{\frac{\omega_{1}\omega_{2}}{4C_{1}C_{1}}}$ in terms  of individual capacitors $C_{1,2}$ and resonator frequencies $\omega_{1,2}= ((L_{1,2}+L_{c_{1},c_{2}})C_{c})^{-1/2}.$  Lumped element LC resonator with resonance frequency $\omega_{r}/2\pi \simeq 8.2$ GHz couples to the flux qubit with interaction strength $g/\omega_{r}\sim 0.1$ in strong coupling regime. When resonance frequency is in the range of $1.0-10$ GHZ, cavity-qubit coupling strength $g/\omega_{r}\sim 1.0$ is addressed in ultrastrong coupling regime of circuit QED\cite{Blais2004,Bourassa2009,abdulmalikov2010,Pol2010,peropadre2013}.

It is possible to tune the strength of the interaction between qubit and privileged mode in ultrastrong regime when disadvantaged mode interacts with the qubit in strong regime\cite{Dereli2012,Gul2016}. Privileged mode, in which coupling energies concentrated, dominates the system and behaves like the scattering center for the cavity fields. Besides this, the interaction of the qubit with the disadvantaged mode and the interaction between the resonators are treated as perturbation. The ability of tuning each resonator in different coupling regimes makes our model suitable to control the storage and transmission of the photons in various input/output ports.

Employing the resonator normal modes, our coupled cavity system which plays the role of the resonators is described by the so-called effective single privileged mode transformation \cite{Dereli2012,O'Brien1972,O'Brien1983} and  described as
\begin{eqnarray}
H= H_{eff}+H'_{ph}+H_{int},
\end{eqnarray}
where effective hamiltonian is
\begin{eqnarray}
H_{eff}&=&\frac{\omega}{2}\sigma_z+\omega_{eff}[\alpha^{\dag}_{1}\alpha_{1}
+k_{eff}(\alpha_{1}+\alpha^{\dag}_{1})\sigma_{x}]
\end{eqnarray} with the effective mode frequency
\begin{eqnarray}
\omega_{eff}&=&\frac{\omega_{1}k^{2}_{1}+\omega_{2}k^{2}_{2}}{k_{eff}},
\end{eqnarray}in terms of individual resonator frequencies $\omega_{1,2}$ and coupling strength of the qubit-resonator interaction is
\begin{eqnarray}
k^{2}_{eff}&=&k^{2}_{1}+k^{2}_{2}.
\end{eqnarray}
Hamiltonian of the disadvantaged effective
mode is given by
\begin{eqnarray}
H'_{ph}&=&\omega'\alpha^{\dag}\alpha
\end{eqnarray}

with the disadvantaged mode frequency
\begin{eqnarray}
\omega'&=&\frac{\omega_{1}k^{2}_{2}+\omega_{2}k^{2}_{1}}{k_{eff}},
\end{eqnarray}
Interaction between the effective and disadvantaged mode
is described by
\begin{eqnarray}
H_{int}=c_{2}[(\alpha^{\dag}_{1}\alpha_{2}+\alpha_{1}\alpha^{\dag}_{2})
+k_{eff}(\alpha^{\dag}_{2}+\alpha_{2})\sigma_{x}],
\end{eqnarray}
and the strength of the coupling between privileged mode and disadvantaged mode is given by
\begin{eqnarray}
c_{2}&=&\frac{\Delta k_{1}k_{2}}{k^{2}_{eff}},
\end{eqnarray}
where the frequency difference $\Delta=\omega_{1}-\omega_{2}$ is used to control
the perturbative interactions on the effective single-mode model.

Our protocol consists of the storage and transmission of the cavity fields in coupled cavity system. It can be further connected to microwave waveguide or transmission line resonators serving as input/ouput ports for detection and measurement using photon statistics. In Fig.1.b. we use artificial atom in $\Lambda$-configuration with three states $|g\rangle$, $|s\rangle$, and $|e \rangle$, from the lowest energy levels to highest as a generic setup of storage and transmission processes for single photon transistor \cite{Chang2007,Manzoni2014,Kyriienko2016}. In our scheme, the excitation of signal, control and gate fields are described by $H_{eff}$, $H'$ and $H_{int}$ respectively.  The resonant interaction between cavity mode $\alpha_{1}$ and $|s\rangle-|e\rangle$ transition and is tuned by coupling strength $\omega_{eff}k_{eff}$. Similarly the interaction between $\alpha_{2}$ mode and $|g\rangle-|e\rangle$ transition tuned by $c_{2}k_{eff}$. Effective JT hamiltonian reveals the inter cavity interactions between lowest lying states $|s\rangle$ and $|g\rangle$ tuned by the $c_{2}=J$ in the presence of the adiabatic elimination of the upper excited states by stimulated Raman transitions.

In Fig.1.c, we choose the privileged mode $\alpha_{1}$ as the first cavity mode and the interaction with the second cavity mode $\alpha_{2}$ is treates as the perturbation. Both transmission (dashed blue row)and reflection (dashed red row) of the cavity fields are tuned by the hopping parameter $J$ in different coupling regimes.
Two microwave waveguide resonators are used to observe photon statistics in input/output ports.  The beam-splitter like interaction hamiltonian $H_{I}$ makes it possible to get resonant absorption and emission when there is a resonance between the $\alpha_{2}$ excitation and the energy of the $n^{th }$ eigenstate of JT-center $H_{eff}$ in the absence of time dependent auxiliary driving field.

Different from the generic models given in Refs. \cite{Chang2007,Manzoni2014,Kyriienko2016} our system uses the hopping parameter of coupled cavity system in controlling lower lying level atomic transitions. Control of the atomic excitations between lower lying states is governed by the photonic nonlinearities rather than external driving fields.

\subsection{Input-Output Theory and Effective Model}
Quantum coherence suffers from dephasing of the artificial superconducting atoms, due to the fluctuations of the cavities which cause phase shift in qubit state depending on the coupling strength. Suppression of dephasing requires dealing with vacuum fluctuations and excitations in coupled qubit-cavity systems \cite{Bertet2005-1,Bertet2005-2,Reuther2011}.

In the input-output theory of cavity QED, output photon flux is proportional to the average number of cavity photons $\langle E^{-}(t)E^{+}(t)\rangle$ in terms of the positive (negative) electric fields $E^{+}(t)(E^{-}(t))$ \cite{Ridolfo2012,Garziano2015}. Similarly, output voltages which are proportional to the electric fields are used in circuit QED. In the presence of vacuum input for a system in its ground state, the destruction operator for the multi-output cavities is expressed as
\begin{eqnarray}
a_{out,i}&=&a_{in,i}+\sqrt{\gamma_{0}}X^{+}_{i}
\end{eqnarray}
where $i=1,2$ is the cavity output port number, $\gamma_{0}$ is dissipation term,$(X^{+}_{i})=\langle j|(a_{i}+a^{\dag}_{i})|k\rangle|j\rangle\langle k|$, and $X^{-}_{i}=(X^{+}_{i})^{\dag}$ corresponds to the higher and lower frequency electric field.  When the input is in the vacuum state, the output cavity photon rate for single mode is given as
\begin{eqnarray}
\Phi_{out,i}&=&\gamma_{0}\langle X^{-}_{i}X^{+}_{i}\rangle
\end{eqnarray}in terms of the port number $i=1,2$. In three-level artificial atoms, spontaneous emission of the cavity photon pairs can be detected by $\Phi_{out,i}$ for each port $i$. These spontaneous virtual photons can be converted into real photons. In our atomic configuration, the transition $|s\rangle\rightarrow|g\rangle$ leads to the spontaneous emission of photons which can leave the cavities.

Apart from the ports $i=1,2$ each carrying fields $X^{+}_{i}=\frac{1}{2}(a_{i}+a^{\dag}_{i})$, we introduce the third port
\begin{eqnarray}
X^{+}_{3,i}=\alpha_{i}+\alpha^{\dag}_{i}
\end{eqnarray} as a linear combination of  the quadrature operators $X^{+}_{3,1}=X^{+}_{1}+X^{+}_{2}=\alpha_{1}+\alpha^{\dag}_{1}$ and $X^{+}_{3,2}=X^{+}_{1}-X^{+}_{2}=\alpha_{2}+\alpha^{\dag}_{2}$. We used the transformation
\[
\begin{pmatrix}
\alpha_{1}\\
\alpha_{2}
\end{pmatrix} = \frac{1}{2}
\begin{pmatrix}
1&1\\
1&-1
\end{pmatrix}
\begin{pmatrix}
a_1 \\
a_{2}
\end{pmatrix}
\] to obtain the cavity modes $a_{1,2}$ in terms of the normal modes $\alpha_{1,2}$. This makes it possible to express the mixture of cavity fields in terms of pure normal modes $\alpha_{1,2}$ in definite port numbers. Interactions between qubit and cavities led to tuning of the parameters to satisfy robustness of the system against the dephasing in terms of the expectation values of cavity fields $a_{1,2}$ and qubit operator $\sigma_{x}$. Using the commutation relations\cite{O'Brien1972,O'Brien1983}
\begin{eqnarray}
 [H,(a^{\dag}_{1,2}-a_{1,2})]&=\omega_{1,2}(a^{\dag}_{1,2}+a_{1,2})+2k_{1,2}\omega_{1,2}\sigma_{x}
\end{eqnarray}
\begin{eqnarray}
 [H,(a^{\dag}_{1,2}+a_{1,2})]&=\omega_{1,2}(a^{\dag}_{1,2}+a_{1,2})
\end{eqnarray}and employing the normal modes $\alpha_{1,2}$, we obtain the relation between the quadrature operator and qubit excitations which is described by
\begin{eqnarray}
 \langle n|\alpha^{\dag}_{1}+ \alpha_{1}|n'\rangle[1-\frac{(E_{n}-E'_{n})}{w_{eff}}]&=&
 -2k_{eff}\langle n|\sigma_{x}|n'\rangle \label{eq:1}.
\end{eqnarray}where $E_{n}$ and $E'_{n}$ are the eigenergies. In Eqn.\eqref{eq:1}, we consider an undamped qubit that is linearly coupled with damped cavities. The cavities are further coupled to a bath at temperature T. These temporal fluctuations of photon number in cavity led to the qubit frequency shift. In the presence of vacuum and ground state excitation of the cavity fields, dephasing of the qubit is described by the two time correlation function $C(\tau)=\langle(\hat\alpha_1(\tau)+\hat\alpha^{\dag}_1(\tau))(\hat\alpha_1(0)+\hat\alpha^{\dag}_1(0))\rangle$ \cite{Bertet2005-1,Bertet2005-2,Reuther2011}.

In our model, expectation values of cavity output field $(\alpha^{\dag}_{1}+ \alpha_{1})$ could be decoupled from the qubit operator $\sigma_{x}$ in definite conditions.
Whenever $E_{n}-E'_{n}= w_{eff}$, qubit state is decoupled, $\langle n|\sigma_{x}|n'\rangle=0$, and perfect transmission is obtained when qubit state is in metastable state $|s\rangle$
corresponding to impedance matching of the cavity field and qubit transitions. Moreover, when $E_{n}=E'_{n}$, the expectation value of privileged mode $\alpha_{1}$ and qubit operator $\sigma_{x}$ can be expressed as
\begin{eqnarray}
 \langle n|\eta+\eta^{\dag}|n'\rangle&=&0
\end{eqnarray} where the hybrid operators,
$\eta= \alpha_{1}+k_{eff}\sigma_{x}$, $\eta^{\dag}= \alpha_{1}^{\dag}+k_{eff}\sigma_{x}$ are the bosonic creation and annihilation operators obeying the commutation relation $[\eta,\eta^{\dag}]=1$. In a frame rotating with the qubit frequency, we describe our JT-system hamiltonian $H$ as
\begin{eqnarray}
H&=& H_{eff}+H_{ph}+H_{int},
\end{eqnarray}
where hamiltonian describing the JT-center is
\begin{eqnarray}
H_{eff}&=& \omega_{eff}\eta^{\dag}\eta-\omega_{eff}k_{eff}^{2}\sigma_{x},
\end{eqnarray}and the $\alpha_{2}$ excitation is
\begin{eqnarray}
H_{ph}&=& \omega'\alpha^{\dag}_{2}\alpha_{2}.
\end{eqnarray}
Interaction between the $\eta$  and $\alpha_{2}$ excitation is written as
\begin{eqnarray}
H_{int}&=& c_{2}(\alpha_{2}^{\dag}\eta+\eta^{\dag}\alpha_{2}).
\end{eqnarray}with the coupling strength $c_{2}=J$.

Thus, our effective model describes coupled cavity-qubit system in terms of coupled bright and dark polaritons, $\eta$ and $\alpha_{2}$ respectively\cite{Dereli2012}. We further decouple the system hamiltonian in terms of upper and lower polaritons. For this purpose, using the Bogoulibov De-Gennes transformation to diagonalize our system, we obtain
\begin{eqnarray}
H= \sum_{l=1,2}E_{l}p^{\dag}_{l}p_{l}
\end{eqnarray} where $p_{l}$ and $p_{l}^{\dag}$ are the annihilation and creation operators
of the polaritons.

The energies are
\begin{eqnarray}
E_{1,2}=\frac{1}{2}[(\omega_{eff}+\omega')\pm\sqrt{(\omega_{eff}-\omega')^{2}+4c_{2}^{2}}]
\end{eqnarray}and the polariton mode operators is described as
\begin{eqnarray}
p_{1}&=&\cos\theta \eta+ \sin\theta \alpha_{2}\\
p_{2}&=&-\sin\theta \eta+ \cos\theta \alpha_{2}
\end{eqnarray}with

\begin{eqnarray}
\tan2\theta=\frac{c_{2}}{\omega_{eff}-\omega'}
\end{eqnarray}where $2\theta$ is the mixing angle.

Comparing with the generic models in circuit QED, our model differs in cavity particle interpretation and output field detection. In cascaded atomic configurations $\Xi$  \cite{Manzoni2014,Kyriienko2016}, dark states results from interference effects between vacuum and ground state. These coherent superpositions of vacuum and multiphoton states are decoupled from higher energy levels by adiabatic Raman transitions and leads to the generation of virtual two photon excitations between atomic levels \cite{Garziano2013,Stassi2013,Huang2014,Garziano2015}.  We use $\Lambda$ structured atomic configuration together with the coupled cavity system and describe the operators $\eta= \alpha_{1}+\sigma_{x}$ and $\alpha_{2}$ as the bright and dark polaritons interacting with coupling strength $c_{2}$. Instead of using coupled bright and dark polaritons, we further decouple the system hamiltonian in terms of upper and lower polaritons to deal with the second order coherence functions for transmission and reflection. To investigate the conversion from virtual to real photons, we employed the expectation values of the quadrature operators $X^{+}_{1,2}$ as the evidence of two photon emission.

\section{Results} \label{sec:results}
For simulation of our single photon transistor, we use the parameters of two-frequency JT system as a test bed in ref\cite{Dereli2012,Gul2016}. Coupling regimes is considered in weak, strong and ultrastrong regimes of circuit QED. Dissipation and dephasing terms are used to describe Non-Equilibrum dynamics of coupled cavities interacting simultaneously with the flux qubit.
\subsection{Power Spectrum and Impedance Matching }
Two time first order correlation function for the electric field operator in terms of privileged mode $\hat\alpha_1+\hat\alpha^{\dag}_1$ is used in power spectrum calculation which is described by
\begin{eqnarray}
P(\omega)=\int_{-\infty}^{\infty}\langle(\hat\alpha_1(t)+\hat\alpha^{\dag}_1(t))(\hat\alpha_1(0)+\hat\alpha^{\dag}_1(0))\rangle
e^{-i\omega t}.
\end{eqnarray} Bloch-Redfield quantum  master equation in Born-Markov approximation is employed to investigate the dissipative dynamic of our system.
Open system dynamics is described by
\begin{eqnarray}
\frac{d\rho}{dt}=-i[H,\rho]+{\cal L}\rho,
\end{eqnarray}
where the Liouvillian superoperator ${\cal L}$ is given by
\begin{eqnarray}
 {\cal L}\rho&=&\sum_{j=1,2}(1+n_{th})\kappa{\cal D}[\hat\alpha_j]\rho+n_{th}\kappa{\cal D}[\hat\alpha_j^\dag]\rho\nonumber\\
 &+&\gamma{\cal D}[\sigma]\rho+\frac{\gamma_\phi}{2}{\cal D}[\sigma_z]\rho,
 \end{eqnarray}
 representing the  average thermal photon number is represented with $n_{th}$. The thermal
occupation number $n_{th}=0.15$ is taken as $100$~mK \cite{Dereli2012,Gul2016}.
${\cal D}$ denotes the Lindblad type damping superoperators,
$\kappa$ shows the cavity photon loss rate. Qubit relaxation and
dephasing rates represented with $\gamma$ and $\gamma_\phi$ , respectively. Resonator decay parameters
are $\kappa_{1}=\kappa_{2}=0.001$ and qubit relaxation and dephasing
parameters are represented with $\gamma=0.001, \gamma_{\phi}=0.01$.

Impedance matching plays an important role in storage and transmission/reflection step of the transistor of $\Lambda$ structured atomic transitions. Among the different sequence of transitions between atomic energy levels $|g\rangle$, $|s\rangle$, $|e\rangle$ \cite{Inamato2014,Koshino2013}, impedance matching occurs in the sequence of the transitions $|g\rangle\rightarrow|s\rangle \rightarrow |e\rangle\rightarrow|g\rangle$ in which $|g\rangle\rightarrow|s\rangle$ transition leads to capture of single photon entering to the coupled cavity system. For the purpose of the controlling transmission and reflection of the incident photon, we employed the Raman transitions by adiabatic elimination of the upper energy level $|e\rangle$. Then, our system, as a coupled cavity model, describe the atomic transitions by the interactions of the excitations lying on the eigenstates of $H_{eff}$ and  $H'$ coupled by the interaction Hamiltonian $H_{int}$.
\begin{figure}[h]
\begin{center}
\subfigure[\hspace{0.001cm}]{\label{fig:1a}
\includegraphics[width=0.4\textwidth]{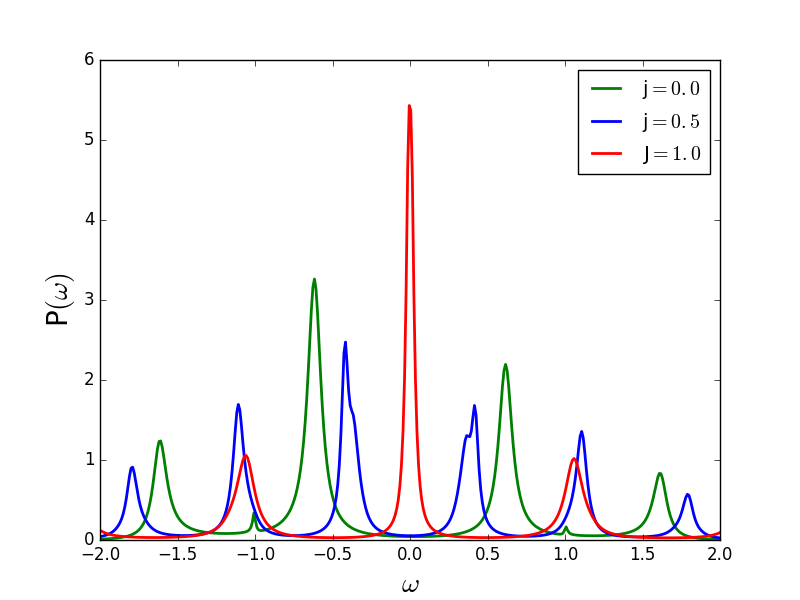}}
\subfigure[\hspace{0.001cm}]{\label{fig:1b}
\includegraphics[width=0.4\textwidth]{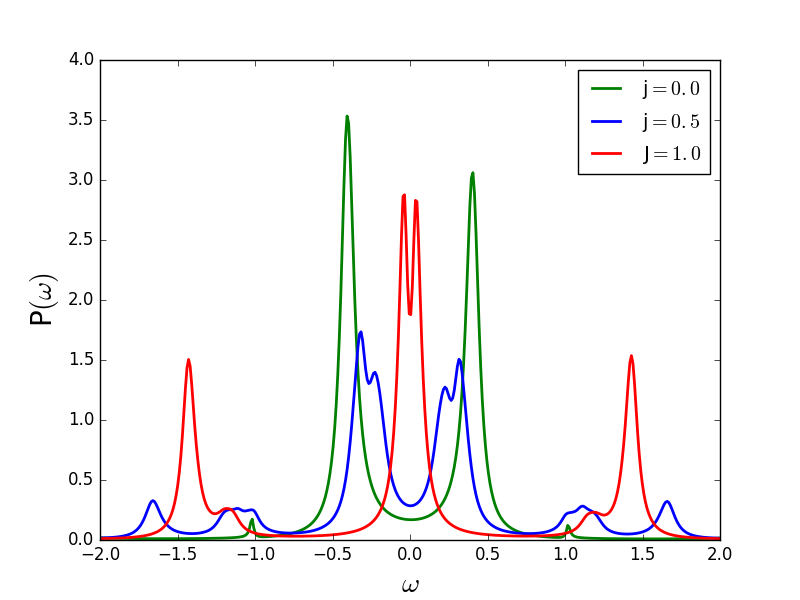}}
\caption{\label{fig3} (Color online) Power spectrum of inter-cavity output photons in uncoupled $J=0.0$ and coupled cavities,$J=0.5, J=1.0$. (a) At intermediate coupling
regime $k=0.5/\sqrt{2}$, central peak becomes higher in amplitude due to increase in hopping.(b)The distance between the two Rabi peaks gets closer in uncoupled scheme.  Rabi Splitting  occurs in coupled cavities in ultrastrong coupling regime $k=1.0/\sqrt{2}$} 
\end{center}
\end{figure}

Effect of cavity interaction $H_{int}$ on  $\alpha_{1}$ via hopping parameter $J$ is shown by the power spectrum in impedance matched condition. Fig.$2$, shows the power spectrum of the extracavity output field in terms of the electric field operator for hopping parameter $J= 0.0$, $J= 0.5$ and $J= 1.0$.  Fig.$2(a)$ shows asymmetric Lorentzian line shape of the Rabi peaks in weak  coupling regime $k=0.5/\sqrt{2}$. When the hoping paremeter reach $J=1.0$,
the two  Rabi sidebands become symmetric in amplitude and localized equally in distance from the central peak.
In Fig.$2(b)$ the distance between the asymmetric Rabi peaks  become narrower in the weak coupling regime $k=1.0/\sqrt{2}$ for hopping parameter $J=0.0$ and $J=0.5$. Whereas, symmetric Rabi peaks are localized further away in ultrastrong regime for $J=1.0$. Multiphoton transitions lead to the splitting of the Rabi peaks in coupled cavity scheme.

\begin{figure}[!hbt]
\begin{center}
\subfigure[\hspace{0.015cm}]{\label{fig:2a}
\includegraphics[width=0.4\textwidth]{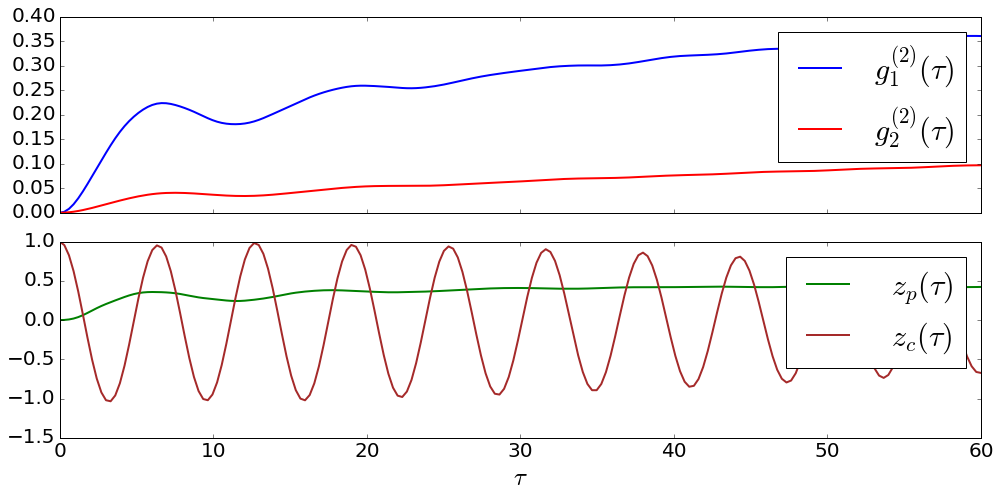}}
\subfigure[\hspace{0.015cm}]{\label{fig:2b}
\includegraphics[width=0.4\textwidth]{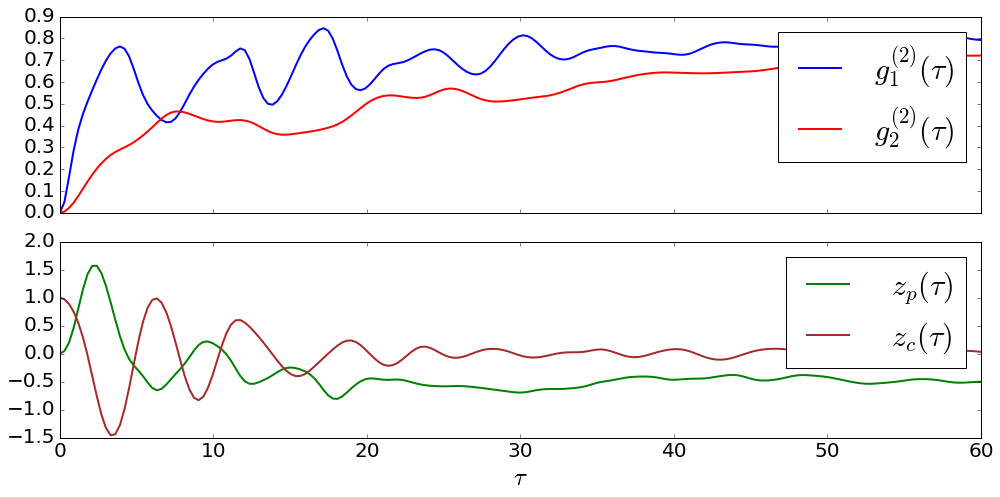}}
\subfigure[\hspace{0.015cm}]{\label{fig:2c}
\includegraphics[width=0.4\textwidth]{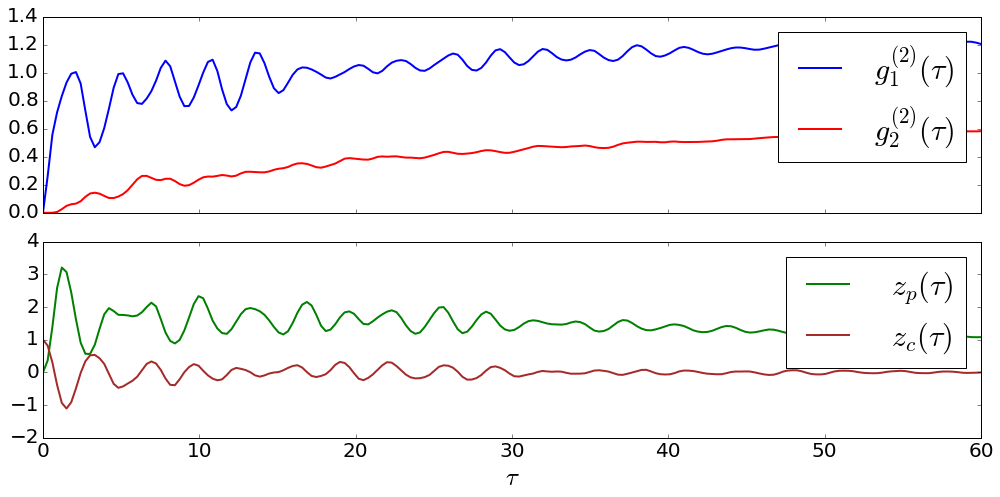}}
\caption{\label{fig3} (Color online) Top panel shows temporal behavior of second order coherence functions for upper  and lower  polaritons in weak, strong and ultrastrong coupling regimes,(a),(b), (c) respectively. $\tau$  is the time-delay between two measurements and defined in units of $\mu s$. Photon antibunching (bunching) occurs when $g^{(2)}_{i}(\tau)>g^{(2)}_{i}(0) (g^{(2)}_{i}(\tau)<g^{(2)}_{i}(0) )$ where $i$ represents output cavity ports. The value of $g^{(2)}(0)\ll 1$ corresponds to antibunching, and
used as the signature of photon blockade. Population imbalances of polariton and cavity fields are presented in lower panel.
(a)Both upper (blue) and lower(red) polaritons are in blockade regime and antibunching. Coherent oscillations are seen in cavity population imbalance and upper polariton get higher populated in the system.(b) Both polaritons are antibunching. Lower polaritons becomes higher populated in the system.
(c)Upper polaritons are higher populated than the lower polaritons.
}
\end{center}
\end{figure}

\subsection{Photon Blockade and Input/Output Theory }
The photon blockade appears as the underlying mechanism of realizing the correlated photons in both uncoupled and coupled cavity systems. Photon blockade makes transmission of an another identical photon conditioned on the presence or absence of previous photon residing in the cavity. Second photon can not be excited in the presence of another one in the cavity. The competition between photon blockade and photon
hopping in coupled cavity systems results in localized and delocalized regimes.
To illustrate the photon blockade, we use the normalized second order coherence functions which are  described as
\begin{eqnarray}
g_{i}^{(2)}=\frac{\langle O^{\dag}_{i}(t)O^{\dag}_{i}(t+\tau)O_{i}(t)O_{i}(t+\tau)\rangle}{\langle O_{i}^{\dag}(t)O_{i}(t)\rangle^{2}}
\end{eqnarray}
where $i=1,2$ are used in place of the cavity field operators. For $\theta=\frac{\pi}{4}$, polariton states are described as the normal modes of $\eta$ and $\alpha_{2}$
\[
\begin{pmatrix}
p_{1}\\
p_{2}
\end{pmatrix} = \frac{1}{\sqrt{2}}
\begin{pmatrix}
1&1\\
-1&1
\end{pmatrix}
\begin{pmatrix}
\eta \\
\alpha_{2}
\end{pmatrix}.
\]

Population imbalances of cavity field (polariton) is defined as $z_{c,p}(t)=(n_{1}-n_{2})/(n_{1}+n_{2})$ where $n_{j}=$Tr$
\hat{\alpha}_{j}^{\dag}\hat{\alpha}_{j}\hat{\rho}$  for $j=1,2$
corresponds to the two cavity (polariton) number operators. The total photon (polariton) number is given by $N=n_{1}+n_{2}$. In our calculations, first cavity contains single photon initially.

In the strong and ultrastrong coupling regimes, the time elapsed by the photon inside cavity gets larger than the decay of the qubit. Then, the photon decay leads to
an anti-bunched train of photons leaving the cavity output ports. Due to the asymmetrical couplings of qubit with privileged and disadvantaged modes $\alpha_{1,2}$, we use the population imbalance between coupled cavities and  bunching and antibunching behavior of photons in coupled cavity system.
In this manner, transmission and reflection of the cavity field in single field transistor can be studied by the delocalization-localization transition in the presence of virtual photons of coupled cavity systems.

In Fig.$3$, the top panel shows the second order correlation functions of upper (blue) and lower component (red) polaritons and the second panel shows population
imbalances of polaritons (green) and cavities (brown) in weak, strong and ultrastrong coupling regimes respectively. In blockade regime, delocalization-localization transitions in definite cavities is tuned by the hopping parameter $J$. Localization in state $|s\rangle$ allows transmission whereas the localization in state $|g\rangle$ cause reflection of the cavity field. The process of capturing the incoming single photon while triggering a spin flip from $|g\rangle$ to $|s\rangle$ give rise to switching of reflection to transmission of cavity output fields. Conversion between the reflected and transmitted field is conditioned on the switching the transitions $|s\rangle\leftrightarrow|g\rangle$. Fig.$3.a$ shows coherence functions starting in a blockade regime, $g^{(2)}_{1,2}(0)\ll1$, in weak coupling regime, $k=0.1/\sqrt{2}$. Since $g^{(2)}_{1,2}(\tau)>g^{(2)}_{1,2}(0)$, both polaritons are antibunching. Coherent oscillations are seen in population
imbalance of cavities and upper polaritons get higher populated.
In the strong coupling regime, $k=0.5/\sqrt{2}$, Fig.$3$.b shows that both of the coherence functions of upper and lower polaritons are anti-bunching. Contrary to the weak regime, lower polaritons get higher populated and cavity fields are in localized regime favored by the strong qubit cavity coupling. In Fig.$3$.c, upper polaritons becomes higher populated in ultrastrong coupling regime, $k=1.0/\sqrt{2}$. Moreover, delocalization-localization transition occurs in cavity fields. The photon hopping resets localization-delocalization transitions by driving the atomic transitions between lowest lying states. Recycling the atomic transitions by virtue of optical nonlinearities in coupled cavities results in repeated cyclic pattern of delocalization-localization transitions in cavity output fields. Resetting transmission conditioned on the cyclic atomic transitions makes our system plausible for single photon transistor architecture.
\begin{figure}[h]
\begin{center}
\subfigure[\hspace{0.001cm}]{\label{fig:1a}
\includegraphics[width=0.4\textwidth]{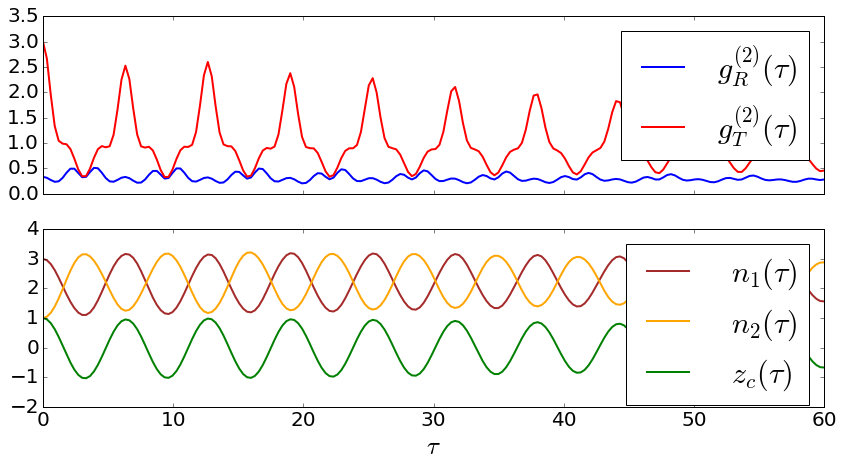}}
\subfigure[\hspace{0.001cm}]{\label{fig:1b}
\includegraphics[width=0.4\textwidth]{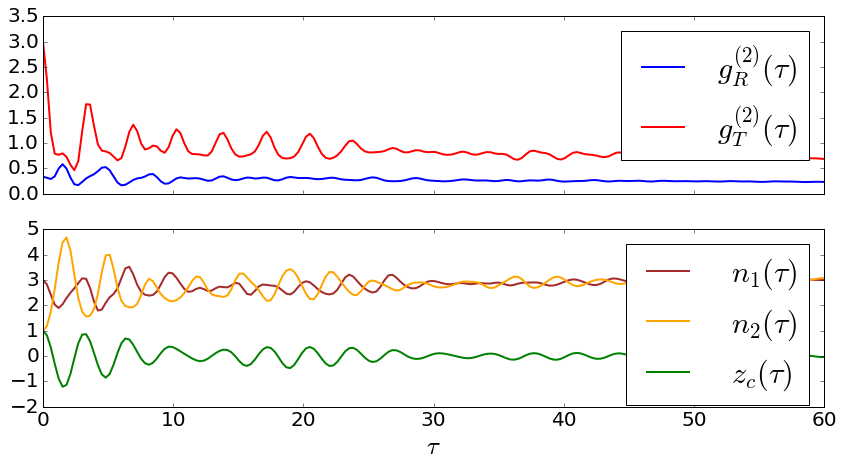}}
\caption{\label{fig3} (Color online)Second order coherence functions (top panel) and expectation values (lower panel) for transmission and reflected output cavity fields in the presence of multilevel transitions. Populations imbalance of coupled cavity system is shown in the lower panel. In weak couping regime, $k=0.1, J=0.5$, bunching of transmitted cavity field and oscillations of multilevel transitions are shown in (a). delocalization-localization transition of cavity fields for the strong coupling regime $k=0.5, J=1.0$ is shown in (b.}
\end{center}
\end{figure}

Input-Output formalism can be extended to the systems containing both single and few photon transport in terms of the output cavity fields. S-matrix elements are used to describe the relation between input output analysis and scattering theory in both single and two photon transport \cite{Fan2010}. In the presence of intermodal coupling between two modes of coupled waveguide interacting with a whispering-gallery-atom, second order coherence functions are used to describe the reflected and transmitted output resonator fields \cite{Shi2013}. When the system is composed of ladder-type three level atom coupled to a waveguide, lower and upper atomic transitions are used to control the oscillating reflected and transmitted waveguide output photons \cite{Kolchin2011}. In the presence of vacuum input, standard input output formalism makes it possible to investigate the vacuum induced virtual photons in expectation values of output cavities and correlations functions\cite{Garziano2013,Stassi2013,Huang2014,Garziano2015}.

In Fig.$(4)$, we show the second order correlation functions $g^{(2)}_{R,T}$ (top panel) and the expectation values $n_{1,2}$ (lower panel) of reflected and transmitted cavity output fields $X^{+}_{3,1}$ and $X^{+}_{3,2}$ in weak and strong coupling regimes. The population imbalance of the coupled cavities is investigated in terms of the normal modes $\alpha_{1,2}$ (lower panel). Fig $4.a$, Second order coherence function of bunching transmitted cavity field  $g^{(2)}_{T}(\tau)$ shows both peak and dips in weak coupling regime $k=0.1$ and $J=0.5$. Difference between the amplitudes of the expectation values of output cavity fields $n_{1,2}$ and population imbalance for coupled cavities implies the spontaneous conversion of virtual photons into real photons. Both expectation values of cavity output fields $n_{1,2}$ and cavity normal modes are in oscillation (delocalized) regime tuned by the hopping parameter $J$. Fig.$4$b shows the delocalization-localization transition of $n_{1,2}$ and population imbalance in strong regime, $k=0.5$ and $J=1.0$.

Different from the generic models\cite{Manzoni2014,Kyriienko2016}, spontaneous emergence of synchronization \cite{Pikovsky2001,Gul2016} in the presence of virtual photon conversion \cite{Garziano2013,Stassi2013,Huang2014,Garziano2015} makes our model advantageous in coupled cavity systems for switching and sending photons. When the single photon gate field is sent to the coupled cavity system, the incoming field is split into two fields incident both on the left and right cavities by time-reversed process of
single-photon generation. It follows that the reflected and transmitted fields shows different statistics where the reflected field is purely scattered field due to the artificial atom and the transmitted field consists both the incident and scattered field. This mixture of the transmitted field makes it difficult to determine the detection of the photons in output ports\cite{Chang2007,Manzoni2014,Kyriienko2016}. For this purpose, we employed the third port in coupled cavity system which describes the  transmitted field in terms of the operators $X^{+}_{3,1}=X^{+}_{1}+X^{+}_{2}=\alpha_{1}+\alpha^{\dag}_{1}$ and $X^{+}_{3,2}=X^{+}_{1}-X^{+}_{2}=\alpha_{2}+\alpha^{\dag}_{2}$ as a linear combination of both reflected and transmitted fields. While both of the reflected and transmitted fields $\alpha_{1,2}$ are antibunching in Fig.$3$, the quadrature operators $X^{+}_{3,1}$ and $X^{+}_{3,2}$ which are responsible for the virtual photon conversion makes it possible to observe bunching in photon statistics of output port in first panel of fig $4.a$.

In oscillating (delocalization) regime, the anti-phase synchronized oscillatory dynamics of both reflected and transmitted fields $n_{1,2}(\tau)$ are governed by the lower level atomic transitions $|g\rangle \leftrightarrow |s\rangle$. In the second panel of fig.$4.a$, the $|g\rangle \rightarrow |s\rangle$ transition leads to the transmitted regime by making the initially higher populated cavity with $n_{1}(\tau)$ lower populated while the second cavity with $n_{2}(\tau)$ gets higher populated. Whereas, $|s\rangle \rightarrow |g\rangle$ transition leads to the reflected regime with higher populated $n_{1}(\tau)$  on the first cavity by blocking second photon entrance. Being in accordance with the $|g\rangle \leftrightarrow |s\rangle$ transitions, switching on the reflected regime to transmitted regime indicates the accumulate and firing process \cite{Pikovsky2001} of the single photon transistor. Accumulation (firing) process in reflected (transmitted) regime leads to weak (strong) bunching in coherence function of the transmitted field. Moreover, transition between reflected and transmitted regimes give rise to the sudden enhancement of the bunching in the form of the spikes with peak (dip) values resulting in the constructive (destructive) interference of the reflected and transmitted field in output port. In addition to the single cycle of the spiking, both of the oscillating individual cavity populations $n_{1,2}(\tau)$ and population imbalance $z_{c}(\tau)$ between two cavities becomes phase locked synchronized with the spike train \cite{Dayan2001,Gerstner2002} of coherence function \cite{Gul2016}. In delocalization-localization  transition, there is destruction of phase locked synchronization due to the damping in amplitudes of both spike train and individual cavity populations as signature of the strong coupling regime with $k=0.5$.

In generic single and two-sided cavity system \cite{Manzoni2014,Kyriienko2016}, avalanche of gain photons at the output port relies on the atomic excitations between the upper levels of the atom triggered by the single photon entrance into cavity. Whereas, in our system, gain enhancements relies on the triggering of the spontaneous virtual photon conversion and beam-splitter structure of the cavity fields in output ports. Contrary to the case in which the photon fields are mixture of the reflected and transmitted fields $a_{1,2}$ used for the gain enhancement in the generic models, we used the reflected  and transmitted fields $a_{out,1}$,$a_{out,2}$ that contribute the output port in a beam-splitter structure via operators $X^{+}_{3,1}$ and $X^{+}_{3,2}$. The presence of virtual photon conversion reveals the gain enhancement in population of individual cavities in terms of the linear combinations of the operators $X^{+}_{1,2}$ with $n_{1,2}(\tau)$ in fig $4.a$. In the presence of multilevel transitions, branching ratio of decay rates $\Gamma_{|e\rangle \rightarrow |g\rangle}$ and $\Gamma_{|e\rangle \rightarrow |s\rangle}$ of the $|e\rangle \rightarrow |g\rangle$ and $|e\rangle \rightarrow |s\rangle$  transitions is given by $n\sim\frac{\Gamma_{|e\rangle \rightarrow |g\rangle}}{\Gamma_{|e\rangle \rightarrow |s\rangle}}$ and used for the description of effective gain \cite{Chang2007}. In our model, we used the ratio $n\sim\frac{n_{2}(\tau)}{n_{1}(\tau)}$ in terms of the transmitted and reflected field populations $n_{2,1}(\tau)$. In fig $4.a$, when the system is in the transmitted regime, $n_{2}(\tau)>n_{1}(\tau)$, $n\sim3$. Similarly, in fig $4.b$, $n\sim4.5$ indicates the gain enhancement in the favor of transmitted field when we go to strong coupling regime.

\section{CONCLUSION}\label{sec:conclusion}
To summarize, we investigated the single photon transistor in the two frequency JT system of the two coupled resonators interacting with a flux qubit simultaneously. Effective single mode transformation is taken into account to describe the system. In input-output relations, extracavity field is analysed in terms of the quadrature operator of the normal modes. Our system hamiltonian is diagonalized via Bogoulibov transformation and population imbalances of both cavity fields and polaritons are employed for the localization-delocalization transition. Conditioned on the atomic transitions tuned by the hopping parameter $J$, reflection and transmission of the cavity output fields are investigated in the presence of virtual photons.

\begin{acknowledgements}

Y. G. gratefully acknowledges support by Bo\u{g}azi\c{c}i University
BAP project no $6942$.  Author thanks \"{O}zg\"{u}r E. M\"{u}stecapl{\i}o\u{g}lu and O. Teoman Turgut for discussion.
\end{acknowledgements}

%

\end{document}